\documentclass[12pt]{article} 

\begin{document}

\newcommand{\vk}{{\vec k}}
\newcommand{\vK}{{\vec K}} 
\newcommand{\vb}{{\vec b}} 
\newcommand{{\vp}}{{\vec p}} 
\newcommand{{\vq}}{{\vec q}} 
\newcommand{\vQ}{{\vec Q}}
\newcommand{\vx}{{\vec x}}
\newcommand{\beq}{\begin{equation}}
\newcommand{\eeq}{\end{equation}} 
\newcommand{\half}{{\textstyle \frac{1}{2}}} 
\newcommand{\gton}{\stackrel{>}{\sim}}
\newcommand{\lton}{\mathrel{\lower.9ex \hbox{$\stackrel{\displaystyle
<}{\sim}$}}} \newcommand{\ee}{\end{equation}}
\newcommand{\ben}{\begin{enumerate}} 
\newcommand{\een}{\end{enumerate}}
\newcommand{\bit}{\begin{itemize}} 
\newcommand{\eit}{\end{itemize}}
\newcommand{\bc}{\begin{center}} 
\newcommand{\ec}{\end{center}}
\newcommand{\bea}{\begin{eqnarray}} 
\newcommand{\eea}{\end{eqnarray}}
\newcommand{\beqar}{\begin{eqnarray}} 
\newcommand{\eeqar}[1]{\label{#1} \end{eqnarray}} 
\newcommand{\pleft}{\stackrel{\leftarrow}{\partial}}
\newcommand{\pright}{\stackrel{\rightarrow}{\partial}}

\begin{center}
{\Large {\bf{Where is the charm quark energy loss at RHIC?}}}

\vspace{1cm}

{ Magdalena Djordjevic and Miklos Gyulassy }

\vspace{.8cm}

{\em Dept. Physics, Columbia University, 538 W 120-th Street,\\ New York,
       NY 10027, USA } 

\vspace{.5cm}

{February 21, 2003}
\end{center}

\vspace{.5cm}

\begin{abstract}
  Heavy quark energy loss in a hot QCD plasma is computed taking into
  account the competing effects due to suppression of zeroth order
  gluon radiation below the plasma frequency and the enhancement of
  gluon radiation due to first order medium induced Bremsstrahlung.
  The results suggest a surprising degree of cancellation between the
  two medium effects for charm quarks and provides a possible
  explanation for the null effect observed by PHENIX in the prompt
  electron spectrum in $Au+Au$ at 130 AGeV.
\end{abstract}

\section*{Introduction}

Recent observations~\cite{Adcox:2002pe}-\cite{qm01} of large
suppression moderate $p_\perp \sim 5$ GeV hadrons produced in $Au+Au$
at the Relativistic Heavy Ion Collider (RHIC) have been interpreted as
evidence for jet quenching of light quark and gluon jets
~\cite{TOMO}-\cite{Wang}. Jet quenching was
predicted~\cite{Gyulassy:1990bh,Gyulassy:1991xb} to occur due to {
  medium induced} radiative energy loss of high energy partons
propagating through ultra-dense QCD matter.  The quenching pattern
therefore provides a novel tomographic tool that can be used to map
out the evolution of the plasma density.  Of course, even without
final state interactions, gluon radiation associated with hard QCD
processes softens considerably the lowest order jet spectra.  This is
taken into account through the $Q^2$ (DGLAP) evolution of the hadronic
fragmentation functions. Medium induced radiation is the extra gluon
radiation that arises from higher twist final state interactions and
depends on the optical thickness or opacity $L/\lambda$ of the medium.

A similar quenching pattern was predicted
~\cite{Shuryak:1996gc,Mustafa:1997pm,Lin:1998bd} to occur for heavy
quark ($c$ or $b$) jets. However, in ~\cite{Dokshitzer:2001zm} it was
pointed out that the heavy quark mass leads to a ``dead cone'' effect
for $\theta<M/E$ that reduces the induced radiative energy loss of
heavy quarks as compared to light partons.  Numerical estimates
indicated that the quenching of charm quarks may be approximately
about a half that of light quarks.  Experimentally, PHENIX
data~\cite{Adcox:2002cg} on ``prompt'' single electron production in
$Au + Au$ collisions at $\sqrt{s}=130$ AGeV have provided a first test
of heavy quark energy loss.  Remarkably, no indication for a QCD
medium effect was found within the admittedly large experimental
errors (see also ref.~\cite{Batsouli_Gyulassy}). In future runs, much
higher statistics and wider $p_T$ range will become accessible.

In this letter we investigate whether the apparent null effect
observed for heavy quark energy loss via single electrons could be due
to other medium effects that compete with the induced radiation.  Our
title is motivated by a similar puzzling null effect seen previously
with light quarks at much lower SPS
energies~\cite{wangsps,Gyulassy:1998nc}.

We study here the non-abelian analog of the
Ter-Mikayelian~\cite{TM1}-\cite{TM2} effect. In
\cite{Dokshitzer:2001zm} the suppression of radiation below a plasma
frequency cutoff was estimated to be only $\sim 10\% $ effect on the
induced energy loss.  However, the Ter-Mikayelian effect on the {\em
  zeroth order} in opacity $(L/\lambda)^0$ radiation has not yet been
considered quantitatively.  The first estimates of the influence of a
plasma frequency cutoff in QCD plasmas were reported in
ref.~\cite{Kampfer_Pavlenko} using a constant plasmon mass
$\omega_{0}$~\cite{Biro_Levai_Muller}-\cite{Levai_Heinz} .  The $k$
dependence of the gluon self energies and the magnitude of
longitudinal radiation were not investigated. In this letter, we
extend those results by taking both longitudinal as well as transverse
modes consistently into account via the frequency and wavenumber
dependent hard thermal (1-loop HTL) self energy
$\Pi^{\mu\nu}(\omega,k)$
~\cite{Kalashnikov:cy}-\cite{Gyulassy_Selikhov}.

The dielectric properties of an isotropic  plasma
lead to a transverse gluon self energy $\Pi_T(\omega, \vk)$ 
with $\Pi_T(\omega_{pl}(0),0)=\omega_{pl}^2(0)\approx \mu^{2}/3$, 
where $\mu\approx gT$ is the Debye screening mass of a plasma at temperature
$T$ in the deconfined phase. In addition, 
long wavelength collective longitudinal gluon modes arise with $\Pi_L(\omega_{pl}(0),0)=
\omega_{pl}^2(0)$. This dynamical gluon mass
suppresses the radiation of soft $\omega<\omega_{pl}(\vk)$ gluons
and shields the collinear $k_\perp\rightarrow 0$ singularities 
that arise for massless quarks.

The second part of our study is to generalize the GLV opacity
expansion method~\cite{GLV} to compute the first order induced energy
loss including the kinematic effects due to both heavy quarks and
massive transverse gluons. These mass effects change the formation
times of the gluons and hence the destructive (LPM) interference
pattern that reduces the radiative energy loss in comparison to the
incoherent Bethe-Heitler limit. For ultra-relativistic light partons,
the formation times are generally long compared to nuclear dimensions.
The energy loss $\Delta E^{ind}\propto L^2$ for very high energy light
partons depends quadratically on the thickness of a static QCD
medium~\cite{BDMS,ELOSS,GLV}, while for heavy quarks we show below
that it is closer to a linear $\propto L^1$ dependence.
 
In this letter, we answer the question in the title by showing that
there is a surprising degree of cancellation between the reduction of
the zeroth order energy loss due to the QCD Ter-Mikayelian effect and
the induced radiative GLV energy loss for heavy quarks.  The technical
details of the derivations and further discussion will be presented
elsewhere~\cite{Djordjevic_Gyulassy_TM,Gyulassy_Djordjevic}.  Our main
result is that while the dead cone effect~\cite{Dokshitzer:2001zm} is
not sufficient by itself to explain the null effect by
PHENIX~\cite{Adcox:2002cg}, the approximate cancellation of medium
effects for heavy quarks could be.

\section*{Plasmon Effect on 0$^{th}$ order Energy Loss}

Our first goal was to test the accuracy of the commonly used
simplifying assumption of neglecting the $\vec{k}$ dependence of
effective plasmon mass as in~\cite{TOMO,Kampfer_Pavlenko,GLV} and also
neglecting longitudinal modes.  As shown below and in more detail
in~\cite{Djordjevic_Gyulassy_TM}, this assumption was found to be
accurate to $\sim10\%$ as long as the {\em asymptotic} plasmon
mass~\cite{Rebhan}, $\omega_\infty=\mu/\sqrt{2}$ rather than the long
wavelength $\omega_{pl}(k=0)=\sqrt{2/3} \; \omega_{\infty}$ is
employed. The accuracy of the approximation also improves dramatically
as the mass of the heavy quark increases.

We computed the plasmon effect on the (zeroth order in opacity) vacuum
energy energy loss, $\Delta E^{(0)}_{med}=\int dk (dI_T/dk+dI_L/dk)$
using the optical theorem in the soft gluon limit ignoring the spin of
the heavy quark.  The zeroth order transverse and longitudinal
radiated energy losses per wave number for an initial quark jet with
four momentum $P=(E,\vec{\mathbf{p}})$ are found to be given by

\beqar
\frac{dI_{T}}{d|\vec{\mathbf{k}}|}&=&\frac{C_{F}}{\pi } \;
\frac{4  {\vec{\mathbf{k}}}^{2}
\vec{\mathbf{p}}^{2} \omega _{T}^{2} (\omega _{T}^{2}-\vec{\mathbf{k}}^{2})}
{\omega _{T}^{2}\mu^{2}-(\omega _{T}^{2}-\vec{\mathbf{k}}^{2})^{2}}
\int_{-1}^1 d\cos\theta \; \frac{\alpha _{S}(Q^2)}{(Q^2-M^2)^2} \sin^2\theta
 \nonumber \\ 
\frac{dI_{L}}{d|\vec{\mathbf{k}}|}&=&\frac{C_{F}}{\pi }
\frac{4 \vec{ \mathbf{k}}^{2}
\vec{\mathbf{p}}^{2}(\omega _{L}^{2}-\vec{\mathbf{k}}^{2})}
{\mu^{2}-(\omega _{L}^{2}-\vec{\mathbf{k}}^{2})}
\int_{-1}^1 d \cos \theta \; \frac{\alpha _{S}(Q^2)}{(Q^2-M^2)^2} 
\left(\cos \theta + \frac{\vec{\mathbf{k}}^{2}}{2 \vec{\mathbf{p}}^2}\right)^2
\eeqar{e1}
where
$Q^2= 2Pk+k^2+M^2$ is the invariant mass of the intermediate state,
  $k^2=\omega^2-\vec{\mathbf{k}}^2$ and
$Pk= E\omega-|\vec{\mathbf{p}}||\vec{\mathbf{k}}| \cos\theta$
with $\omega=\omega_T(k)$ for the case of transverse radiation
and  $\omega =\omega_L(k)$ 
for longitudinal radiation.
The well known dispersion relations, $\omega _{T}(k)$ and $\omega _{L}(k)$ are obtained
from $\omega ^{2}\epsilon_{T}(\omega,k)-{\mathbf{k}}^{2}=0$ and $\epsilon_{L}(\omega,k)=0$ 
respectively. Analytic expressions for $\epsilon_{T}$ and $\epsilon_{L}$ in 
the HTL or equivalently, semi-classical transport linear response 
approximations can be found in refs.~\cite{Pisarski:cs,Rebhan,Gyulassy_Selikhov}. The angular integration can be performed analytically if $\alpha_S$ does 
not run. We perform
the integration with running coupling at the scale $Q^2$ as in conventional
DGLAP evolution.

\begin{center}
\vspace*{5.3cm}
\includegraphics{Charm1.ps}  
\vskip 28pt
\begin{minipage}[t]{15.0cm}
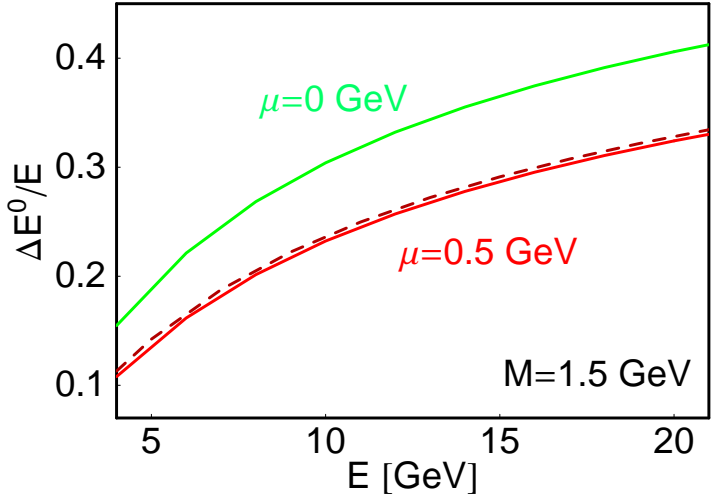
{\small {FIG~1.} The reduction of the zeroth order (vacuum) energy loss for charm quark 
due to the QCD Ter-Mikayelian plasmon effect is shown
as a function of the charm quark energy. 
The upper curve shows the vacuum energy loss if gluons are 
treated as massless and transversely polarized.  The lower solid 
curve shows medium modified (but zeroth order in opacity) 
transverse fractional 
energy loss. The dashed curve shows the negligible additional effect of 
longitudinal plasmons.}
\end{minipage}
\end{center}
\vskip 4truemm

The numerical results for a medium characterized by a Debye screening
scale $\mu=0.5$ GeV in Fig.1 show that the longitudinal plasmon
contribution to the energy loss is indeed negligible for the range of
energies considered.  However, the transverse plasmon mass effect
reduces the zeroth order energy loss by, $ \sim 30 \% $ , relative to
the vacuum case.  The plasmon mass cutoff would effectively {\em
  enhance} the yield of high transverse momentum charm quarks were it
not for the extra medium induced radiation discussed in the next
section.

\section*{\nolinebreak{Medium induced radiative energy loss}}

The second effect of the medium on the the propagation of heavy quarks
is induced radiation caused by the multiple interactions with partons
in the medium. In order to compute medium induced energy loss for
heavy quarks we extended the GLV method~\cite{GLV_1} to include the
kinematic modification of the propagators. The new feature here is
that unlike in previous applications, both the quark and and gluon
plasmon mass ($M$ and $m_{g}$ respectively) are taken into account.

Since we found~\cite{Djordjevic_Gyulassy_TM} that the dielectric
plasmon effect can be well approximated numerically neglecting the
longitudinal modes and using the wavenumber independent asymptotic
mass $\mu/\sqrt{2}$ as the effective gluon mass, we will use this
simplifying assumption in computing the induced radiative energy loss.

The heavy quark generalization of the GLV energy loss to first order
in the opacity $L/\lambda= \int dx \sigma \rho $ is found to
be~\cite{Gyulassy_Djordjevic}:

\medskip 

\beqar
\Delta E^{(1)}_{ind} &=& \frac{C_{F}\alpha_{S}}{\pi} \frac{L}{\lambda}
\int_0^1 dxE \int 
\frac{\mu^2 dq_\perp^2}{(q_\perp^{2}+\mu ^{2})^{2}} \int \frac{ dk_\perp^2 \; 
\theta (x E-k_\perp)}{( 
\frac{4 E x}{L} )^{2} 
+ ({k_\perp}^{2} + M^{2}x^{2} + m_{g}^{2})^{2}} \nonumber \\
&\; & \hspace{-0.5in} \left\{ {k_\perp}^{2} - M^{2}x^{2} - m_{g}^{2} + 
\frac{({q_\perp}^{2} - { k_\perp}^{2} + M^{2}x^{2} + m_{g}^{2})
( {k_\perp}^{2} + M^{2}x^{2} + m_{g}^{2})}
{ \sqrt{((k_\perp-{q_\perp})^{2} + M^{2}x^{2} + m_{g}^{2}) 
(( {k_\perp}+{q_\perp})^{2} + M^{2}x^{2} + m_{g}^{2}) }} \right\},
\eeqar{e2}
where $q_\perp$ is the magnitude of the transverse momentum transfer between a target parton and a 
jet, ${ k_\perp}$ is the magnitude of the transverse momentum of the radiated
gluon, and $\lambda$ is 
the mean free path of the gluon. The simple analytic form of the destructive interference
factor involving $(4 E x/L)^2$ arises
for an assumed exponential distribution $\exp(-\Delta z/L)/L$ between scattering centers.
The expression in $\{ \cdots\}$ reduces in the $M=m_g=0$
limit to $2 k_\perp^2 \theta(q_\perp-k_\perp)$, 
and the GLV asymptotic result eq.(127) of \cite{GLV_1} can be
recovered. We ignore the finite kinematic bounds on
$q_\perp<\sqrt{6ET}$.  The $q_\perp^2$ integral can then be performed analytically 
but is cumbersome. Qualitatively, the effect of increasing $M$ is to reduce
the relevance of the formation time factor $(4 E x/L)^2$  in the denominator
 of the integrand. Formally, setting this factor to zero
recovers the Bethe-Heitler result with $\Delta E^{(1)}_{ind} \propto \alpha_s E L/\lambda
$ modulo a logarithmic factor.

The numerical results for the first order induced radiative energy
loss are shown on Fig.2 for charm and bottom quarks. We fix the
effective static plasma opacity to be $L/\lambda=4$ from the analysis
of light quark quenching in $Au+Au$ at $130$ GeV from the results of
Levai et al~\cite{Levai_opacity}. We assume here that $\alpha_s=0.3,
\; \mu=0.5\;{\rm GeV},$ and $\lambda=1$ fm for the plasma parameters.
In the energy range, $E \sim 5-15$ GeV, the Ter-Mikayelian effect
reduces the induced energy loss in this extension of the GLV approach
somewhat more than in the BDMS approximation\cite{Dokshitzer:2001zm}.
However, on an absolute scale, this only corresponds to a change of
$\delta(\Delta E^{(1)}/E)\sim 0.05$, which is negligible.  Note that
with both dead cone and plasmon mass reduction, there remains a
sizeable induced energy loss fraction $\Delta E^{(1)}/E\approx 0.2$
for charm quarks while only about half that is predicted for bottom.
Therefore, the dead cone effect by itself is not sufficient to explain
the PHENIX heavy quark null effect.
\begin{center}
\vspace*{6.3cm} \includegraphics{Charm2.ps}

\begin{minipage}[t]{15.0cm}
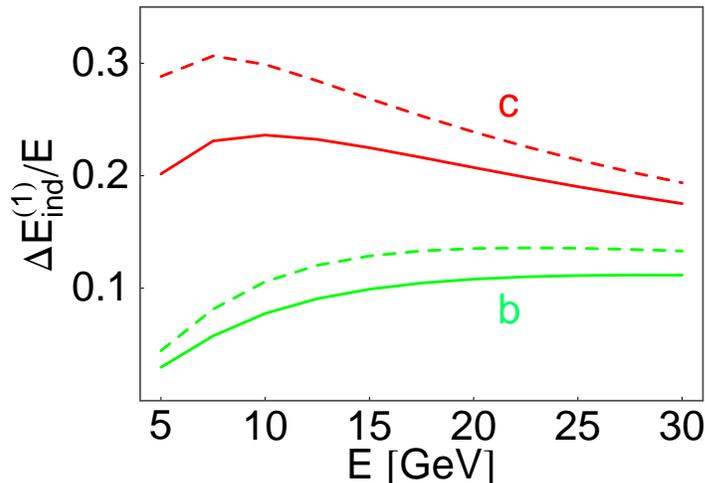
{\small {FIG~2.} The $1^{st}$ order in opacity fractional energy loss for heavy quarks 
with (solid curves) and without (dashed curves) Ter-Mikayelian effect
approximated by a constant $m_g=\mu/\sqrt{2}$ for transverse modes only.
 Upper curves correspond to charm, 
and lower to bottom quarks as a function of their energy in a plasma
characterized by $\alpha_s=0.3, \; \mu=0.5\;{\rm GeV},$ and $L=4\lambda=4$ fm. 
}
\end{minipage}
\end{center}
\vskip 4truemm

\section*{The net heavy quark energy loss}

Fig.3 shows the competition between the medium dependence of the
induced energy loss and the zeroth order energy loss taking into
account the Ter-Mikayelian effect. Even in the absence of a medium
($L=0)$, a charm quark with energy $\sim 10$ GeV suffers an average
energy loss, $\Delta E^{(0)}_{vac}/E\approx 1/3$, due to the sudden
change of the color current when it is formed in the vacuum.  The
dielectric plasmon effect reduces this to about $\Delta
E^{(0)}_{med}/E\approx 1/4$. This contribution is independent of the
thickness of the plasma as long as $L$ is not too small. For very
small $L< 1/m_g$, the plasmon dispersion is smeared out due to the
uncertainty principle, and $\Delta E^{(0)}_{med}$ must approach
$\Delta E^{(0)}_{vac}$ from below.

The induced contribution, $\Delta E^{(1)}_{ind}$, on the other hand, increases with $L$.
 Note that for charm quarks the thickness dependence
is closer to the linear Bethe-Heitler like form, $L^1$,
than the asymptotic energy quadratic form\cite{ELOSS,GLV}.
More importantly, the two competing effects tend to cancel to a large extend 
for an effective thickness in the physically relevant range $L$.
\begin{center}
\vspace*{7.4cm} \includegraphics{Charm3.new.ps}

\begin{minipage}[t]{15.0cm}
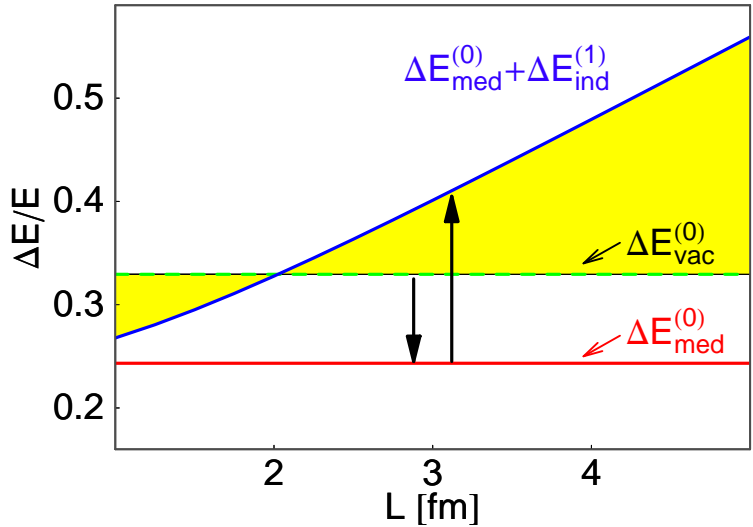
{\small {FIG~3.} The fractional energy loss for a 10 GeV charm quark is plotted 
versus the effective static thickness $L$
of a plasma characterized by $\mu=0.5$ GeV and $\lambda=1$ fm. 
The dashed middle horizontal line corresponds to the
energy loss in the vacuum taking into account the kinematic dead cone
of radiation for heavy quarks~{\protect\cite{Dokshitzer:2001zm}}.
The lower horizontal solid
line shows our estimate of the reduction of the zeroth order 
energy loss due to the QCD analog of the Ter-Mikayelian
effect. The solid curve corresponds to the net
energy loss, $\Delta E^{(1)}_{ind}+\Delta E^{(0)}_{med}$. 
}
\end{minipage}
\end{center}
\vskip 4truemm 

The net energy loss, $\Delta E^{(0)}_{med}+\Delta E^{(1)}_{ind}$ is found to be
remarkably close to the naive vacuum value $\Delta E^{(0)}_{vac}$ for
the effective static opacity medium 
with $L\approx 3$ fm and $\mu\approx 0.5\; {\rm GeV}, \;
\lambda\approx 1$fm).  These results suggest that in addition to the
heavy quark dead cone effect, a high degree of cancellation between
the competing medium effects may be an important factor that could
explain the null effect found experimentally by
PHENIX\cite{Adcox:2002cg} for $Au+Au$ at 130 GeV.  The different
dependence of these effects on the plasma thickness $L$ and on the
transport properties, $\mu$ and $\lambda$, should make it possible to
test this explanation by varying the beam energy, $A$, and centrality
at RHIC and LHC energies.

In closing, we note a third mechanism that could also play a role in
explaining the null effect reported by PHENIX.  As shown in
\cite{ww01}, for quark momenta below $\sim 3T$ detailed balance in a
thermal medium gives rise to a positive feedback due to thermal gluon
absorption that limits the energy loss. For a rapidly expanding
hydrodynamic medium with transverse flow boost rapidity, $\eta_T$,
even large energy loss cannot quench the spectrum below the blue
shifted $p_T\sim 3 T e^{\eta_T}$ thermal distribution {\em if} the
heavy quark comes into local equilibrium. As noted in
\cite{Batsouli_Gyulassy}, for charm quarks there may in fact be only a
small room between the unquenched pQCD spectrum and the fully quenched
hydrodynamical spectrum in the $p_T$ range accessible thusfar at RHIC.
A decisive test of this explanation would be the observation of charm
elliptic flow at RHIC. If, on the other hand, no charm elliptic flow
is observed and the null effect persists to higher $p_T$, then the
cancellation mechanism discussed above
could solve the charm puzzle.\\[2ex]

{\em Acknowledgments:} Valuable discussions with I. Vitev, Z. Lin, J.
Nagle, and W. Zajc on heavy quark production at RHIC are gratefully
acknowledged. This work is supported by the Director, Office of
Science, Office of High Energy and Nuclear Physics, Division of
Nuclear Physics, of the U.S. Department of Energy under Grant No.
DE-FG02-93ER40764.

\end{document}